%% file: main.tex
\documentclass[sigconf]{acmart}

\AtBeginDocument{%
  \providecommand\BibTeX{{%
    \normalfont B\kern-0.5em{\scshape i\kern-0.25em b}\kern-0.8em\TeX}}}

\setcopyright{acmcopyright}
\copyrightyear{2022} 
\acmYear{2022} 
\setcopyright{rightsretained} 
\acmConference[DAC '22]{Proceedings of the 59th ACM/IEEE Design Automation Conference (DAC)}{July 10--14, 2022}{San Francisco, CA, USA}
\acmBooktitle{Proceedings of the 59th ACM/IEEE Design Automation Conference (DAC) (DAC '22), July 10--14, 2022, San Francisco, CA, USA}
\acmDOI{10.1145/3489517.3530420}
\acmISBN{978-1-4503-9142-9/22/07}

\usepackage[normalem]{ulem}

\usepackage{xspace}
\usepackage{multirow}
\usepackage{comment}

\usepackage{amsmath,amsfonts}
\usepackage{algorithm}
\usepackage{graphicx}
\usepackage{textcomp}
\usepackage{xcolor}

\usepackage{algcompatible}

\algblockdefx{FORALLP}{ENDFAP}[1]%
  {\textbf{for all }#1 \textbf{do in parallel}}%
  {\textbf{end for}}

\usepackage{threeparttable}

\usepackage{newfloat}
\DeclareFloatingEnvironment[
    fileext=los,
    listname={List of TableA},
    name=Table A,
    placement=tbhp,
]{tableA}

\newcommand{\serpens}{\textsc{Serpens}\xspace}

\usepackage{tikz}

\newcommand\blfootnote[1]{%
  \begingroup
  \renewcommand\thefootnote{}\footnote{#1}%
  \addtocounter{footnote}{-1}%
  \endgroup
}

\begin{document}
\title[\serpens: A High Bandwidth Memory Based Accelerator for General-Purpose Sparse Matrix-Vector Multiplication]{
\serpens: A High Bandwidth Memory Based Accelerator for General-Purpose Sparse Matrix-Vector Multiplication}

\author[Linghao Song, 
Yuze Chi,
Licheng Guo, and
Jason Cong]{
Linghao Song, 
Yuze Chi,
Licheng Guo, and
Jason Cong}

\email{{linghaosong,chiyuze,lcguo,cong}@cs.ucla.edu}
\affiliation{%
  \institution{University of California, Los Angeles}
  \country{}
}

\begin{abstract}
\input{texfiles/abstract.tex}

\end{abstract}

\maketitle

\thispagestyle{empty}

\section{Introduction}
\label{sec:intro}
\input{texfiles/sec_introduction.tex}

\section{Background and Motivation}
\label{sec:background}

\input{texfiles/sec_background.tex}

\section{\serpens Accelerator}
\label{sec:architecture}
\input{texfiles/sec_architecture.tex}

\section{Evaluation}
\label{sec:evaluation}
\input{texfiles/sec_evaluation.tex}


\section{Conclusion}
\label{sec:conclusion}
\input{texfiles/sec_conclusion.tex}

\bibliographystyle{ACM-Reference-Format}
\bibliography{references}

\end{document}

%% file: texfiles/abstract.tex
Sparse matrix-vector multiplication (SpMV) multiplies a sparse matrix with a dense vector. SpMV plays a crucial role in many applications, from graph analytics to deep learning. The random memory accesses of the sparse matrix make accelerator design challenging. However, high bandwidth memory (HBM) based FPGAs are a good fit for designing accelerators for SpMV. In this paper, we present \serpens, an HBM based accelerator for general-purpose SpMV, which features  memory-centric processing engines and index coalescing to support the efficient processing of arbitrary SpMVs. From the evaluation of twelve large-size matrices, \serpens is $1.91\times$ and $1.76\times$ better in terms of
geomean throughput than the latest accelerators GraphLiLy and Sextans, respectively. We also evaluate 2,519 SuiteSparse matrices, and \serpens achieves $2.10\times$ higher throughput than a K80 GPU. For the energy/bandwidth efficiency, \serpens is $1.71\times$/$1.99\times$,
$1.90\times$/$2.69\times$, and
$6.25\times$/$4.06\times$ better compared with GraphLily, Sextans, and K80, respectively.
After scaling up to 24 HBM channels,
\serpens achieves up to
60.55~GFLOP/s (30,204~MTEPS) and up to 
3.79$\times$ over GraphLily.
The code is available at \url{https://github.com/UCLA-VAST/Serpens}.

%% file: texfiles/sec_introduction.tex
SpMV \blfootnote{This work is supported in part
by the NSF RTML Program (CCF-1937599),
CDSC industrial partners~(\url{https://cdsc.ucla.edu/partners}),
and the Xilinx XACC Program.}
performs the computation of 
$\vec{y} = \alpha\cdot\mathbf{A}\times\vec{x} + \beta\cdot\vec{y}$ where
$\vec{x}$ and $\vec{y}$ are two dense 
vectors, $\mathbf{A}$ is a sparse matrix,
and $\alpha, \beta$ are two scalar constants. SpMV is the core
computation routine in 
a wide range of applications,
such as linear systems solvers~\cite{saad2003iterative}
in scientific computing,
the processing model~\cite{kepner2016mathematical} 
in graph analytics, and
inference of sparse neural networks~\cite{han2015learning}.
In the acceleration of dense algebra,
the tensor size determines the data movement 
and thus researchers can use
an analytic model to coordinate
the computation to achieve very high performance~\cite{wang2021autosa}.
However, it is difficult to accelerate
SpMV because:
(i) 
The vector $\vec{x}$ has only one element at each index that significantly prevents reuse in computation. Thus, it
is hard to achieve a high computation throughout.
(ii)~The irregular distribution 
of non-zeros
in the sparse matrix $\mathbf{A}$ leads to random memory accessing. The memory hierarchy faces high pressure from the random accessing. If we do not optimize the accelerator's memory, the performance will be even lower.

High bandwidth memory (HBM)~\cite{hbmjedec}
exposes more memory channels to users than conventional
DDR memory. HBM-based FPGAs enable accelerator design
and evaluation with high memory bandwidth. It is an
opportunity to accelerate memory-intensive applications
including SpMV with HBM FPGAs. 
GraphLily~\cite{hu2021graphlily} and
Sextans~\cite{song2021sextans} are two 
of the latest
accelerators leveraging HBM FPGAs for sparse workloads.
GraphLily is an FPGA overlay to support graph applications which can be described by an SpMV BLAS
processing model. Sextans is an accelerator for 
sparse matrix-matrix multiplication (SpMM). 
However, there are a few limitations in
existing works for SpMV acceleration:  
(i) GraphLily overlay deploys extra hardware resource to support generalized operations, but
  many of the hardwarelized operations are idle in SpMV.
  So the FPGA hardware is not fully customized
  for SpMV acceleration.
(ii) Sextans has to allocate memory channels to one sparse matrix and two dense matrices in SpMM processing. However, for SpMV acceleration, we can save some memory channels for dense matrices because the dense vector size is smaller than the dense matrix size. Thus, we can allocate more memory channels to speed up the processing
  of the sparse matrix.
(iii) There lacks a modern FPGA-based accelerator for general-purpose SpMV as we treat FPGAs as a competitive candidate to GPUs for computing in data center.


We present \serpens, a high bandwidth memory-based FPGA accelerator for general-purpose SpMV acceleration.
Our features include:
(i) \serpens is an
  HBM FPGA accelerator for general-purpose\footnote{For 'general-purpose' we mean that the accelerator (1) supports a general form SpMV $\vec{y} = \alpha\cdot\mathbf{A}\times\vec{x} + \beta\cdot\vec{y}$ and (2) can run an arbitrary SpMV without re-doing prototyping.} SpMV. \serpens supports arbitrary spares matrices and achieves
  competitive performance to GPUs.
(ii) The memory-centric processing engines (PEs) of \serpens
  enable efficient streaming of sparse matrices. We partition the input dense vector into segments and accumulate output dense vector on chip to process in an output stationary~\cite{sze2017efficient} manner. Thus, we limit the random memory to on-chip BRAMs/URAMs to avoid the high latency of random off-chip memory accessing. The memory-centric PEs also make
  \serpens a scalable architecture.
(iii) \serpens uses an index coalescing to fully utilize on-chip FPGA URAMs to support large-size problems.
  Similar to prior works~\cite{song2021sextans,hu2021graphlily,srivastava2020tensaurus}, we preprocess the spares elements into accelerator-efficient storage.
(iv) We conduct comprehensive evaluations. We evaluate on 12 million-level matrices, \serpens 
  surpasses GraphLiLy by $1.91\times$ and Sextans by $1.76\times$ in terms of geomean throughput. In terms of energy efficiency, \serpens is $1.71\times$ better than GraphLiLy and $1.90\times$ better than Sextans. We evaluate
  on 2,519 SuiteSparse~\cite{davis2011university} matrices, \serpens is $2.10\times$ in terms of throughput, $6.25\times$ in terms of and energy efficiency, and $4.06\times$ in terms of bandwidth efficiency better than a K80 GPU. After scaling up to 24 HBM channels,
\serpens achieves up to
60.55~GFLOP/s (30,204~MTEPS) and up to 
3.79$\times$ over GraphLily.

%% file: texfiles/sec_background.tex
\subsection{High Bandwidth Memory}

Conventional DDR memory provides limited memory bandwidth. 
For example, the DDR4 memory of a 
Xilinx Alveo U250~\cite{u250} 
provides four channels and a total
bandwidth of 77 GB/s. Accelerators for
computation-intensive applications
such as deep learning accelerators~\cite{wang2021autosa}
are able to achieve high performance
with DDR memory. However, in
SpMV and many related
graph applications
the data reuse  is low 
and there are
a large amount of random memory accesses.
Such applications are memory-intensive and require
the support of high memory bandwidth.
HBM based accelerator Xilinx Alveo U280~\cite{u280} 
provides 32 channels and
a total memory bandwidth of 460 GB/s, which
is a good opportunity for the acceleration of 
memory-intensive applications. However, 
it is non-trivial to achieve efficient
HBM channel interconnection~\cite{choi2020hls,choi2021hbm}. Accelerator architects
need to customize their accelerators to fit the HBM
channels to fully reap the bandwidth benefit.

\subsection{SpMV Accelerators on HBM FPGAs}
GraphLily~\cite{hu2021graphlily} and
Sextans~\cite{song2021sextans} are two 
of the latest
accelerators related to \serpens.
They are both accelerators based
on HBM FPGAs.
Although they are
not specialized for SpMV, they are able to
support SpMV processing.

GraphLily~\cite{hu2021graphlily} uses
a BLAS-based processing model~\cite{kepner2016mathematical} which represents graph applications in a generalized SpMV to design
an FPGA overlay as a general accelerator for graph processing. 
To run different graph applications, 
GraphLily configures the data type,
the generalized binary multiplication,
and the generalized reduction. For example, to support a floating-point
SpMV, GraphLily sets the data type to 
float and maps the generalized binary multiplication to arithmetic multiplication and the reduction to 
arithmetic addition. SpMV never uses
the other
hardware instances of the generalized 
operations. Moreover, GraphLily deploys
an arbiter vector unit to load data
from off-chip memory and supply it
to processing engines which have no
bank conflicts. The arbiter vector unit
is flexible in BFS and SSSP.
However, for SpMV
processing, we know the vector accessing sequence in advance. 
Thus, GraphLily does not fully customize
the vector handling
for SpMV.
Nevertheless, the overlay
makes GraphLily support
a wide spectrum of
graph applications
including BFS, SSSP, and PageRank
besides SpMV.

For SpMM acceleration, 
Sextans~\cite{song2021sextans} 
balances the
allocation of memory channels to one sparse
and two dense matrices, 
because SpMM needs to stream on three
large
matrices of
comparable sizes.
Specifically, Sextans allocates 8 channels for the 
sparse matrix and 20 channels for the two
dense matrices. Besides, to achieve
a high computation throughput, Sextans
shares a sparse elements 
with eight
dense matrix elements. The sharing
consumes FPGA logic resource and
on-chip BRAMs.
However, the dense vectors in SpMVs
are quite smaller than the dense matrices
in SpMM. To support an SpMV run,
Sextans configures $N=8$ to run
an SpMM and retires
the first column vector of the SpMM
output as the result
of SpMV.
However, we can allocate less 
channels for the vector and more
channels for the sparse matrix. Moreover,
an SpMV accelerator can save 
the on-chip logic and memory resource 
previously used for the SpMM sharing.

\subsection{Other Related Works}

SpaceA~\cite{xie2021spacea} is a
hybrid memory cube-based
accelerator architecture for SpMV.
Tensaurus~\cite{srivastava2020tensaurus} is an HMC
based accelerator for sparse-dense linear 
algebra.
GraphR~\cite{song2018graphr} utilizes 
an SpMV
processing model for graph acceleration.
However, the three accelerators are evaluated
by simulation rather than real execution.
Fowers et al.~\cite{fowers2014high} designed an FPGA accelerator for SpMV,
but it is on DDR memory and the performance 
is poor. HitGraph~\cite{zhou2019hitgraph} and ThuderGP~\cite{chen2021thundergp}
are FPGA accelerators for graph processing,
but they utilize DDR memory and 
are no specialized for SpMV.

Data format/layout reorganization is a
common technique for boosting
systems performance. For example, CSR5~\cite{liu2015csr5} and HiCOO~\cite{li2018hicoo} are data formats
to accelerate SpMV and sparse tensor processing
on muti-core CPUs and GPUs. TensorFlow~\cite{abadi2016tensorflow} stores data in TFRecord format for fast processing.
For accelerators, \cite{fowers2014high,hu2021graphlily,song2021sextans}, and \cite{srivastava2020tensaurus} all reorganize
data format/layout to be accelerator friendly.

\subsection{Motivation}
The lack of modern HBM-FPGA-based accelerators
for SpMV motivates us to develop \serpens. 
\serpens utilizes massive HBM memory channels
for high-throughput processing of sparse
matrices.
We customize the storing and sharing of
dense vectors 
for the need of SpMV 
to fully utilize FPGA resource.
We architect \serpens as a general-purpose 
accelerator to deliver competitive SpMV
performance to GPUs for data-center computing.

%% file: texfiles/sec_architecture.tex
\subsection{Accelerator Architecture}

\begin{figure*}[tb]
\centering
\includegraphics[width=1.6\columnwidth]{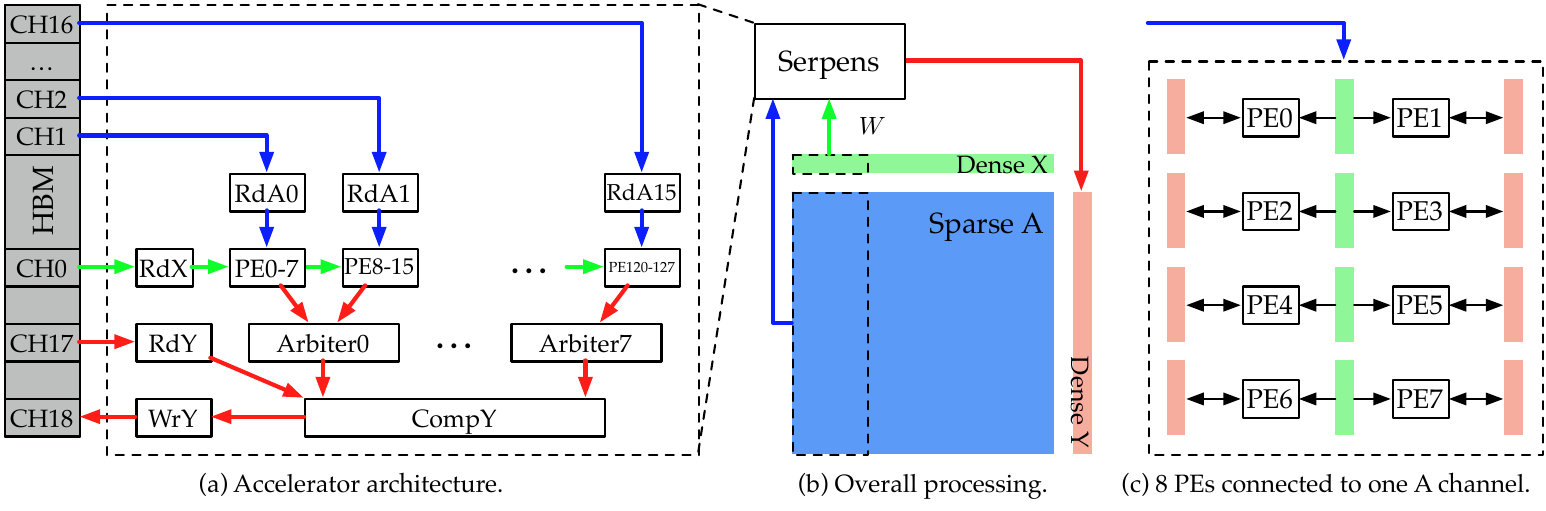}
\vspace{-12pt}
\caption{\serpens accelerator architecture and matrix-vector processing. 
}
\label{figure:overall_arch}
\vspace{-6pt}
\end{figure*}

\subsubsection{HBM Channel Allocation}
Figure\ref{figure:overall_arch} (a) shows the overall architecture of the
\serpens accelerator.
For the off-chip memory accessing in SpMV,
\serpens needs to (1) stream in the sparse
$\mathbf{A}$ matrix, (2) stream in the dense
$\vec{x}$ vector, and (3) stream in the dense
$\vec{y}$ vector and write the result $\vec{x}$ vector. 
The dense vector size is much
smaller than the sparse matrix size.
For example, the matrix {\tt hollywood}
is 1.25 GB while the corresponding
dense vector is 4 MB.
Thus, we allocate one HBM channel for each dense
vector, i.e.,
$\vec{x}$, input $\vec{y}$, and 
output $\vec{y}$, and sixteen HBM channels
for the sparse
$\mathbf{A}$ matrix. In total, \serpens occupies 19 HBM channels and the accumulative
memory bandwidth is 273 GB/s.

\subsubsection{\serpens Modules}
We deploy a read (Rd) or write (Wr) module for each HBM channel. The Rd/Wr modules performs
streaming memory accessing to off-chip HBM.
The bitwidth of the Rd/Wr modules is 512.
For the dense vector Rd/Wr modules, we coalesce
16 floating-point values into a 512-bit segment. For one sparse element, each of the row index, column index, and float attribute 
occupies 32 bits. Because we partition the vectors and matrices in SpMV processing
(Sec.\ref{sec:process_order}), the indices
are limited in a range at each iteration. Thus,
we reduce the index bits and compress a
row-column index pair into 32 bits to save 
memory bandwidth. We encode a sparse element 
with 64 bits. As a result,
for the sparse matrix read module,
we coalesce eight sparse elements into 
a 512-bit segment. 
\serpens deliver the spares elements from
one HBM channel to 8 processing engines (PEs).
One PE performs part of the matrix-vector
multiplication $\mathbf{A}\times\vec{x}$.
We use one arbiter to select computation results from 16 PEs and send the result to 
a CompY module. The CompY module performs the
element-wise ($\alpha,\beta$) multiplications and additions
to obtain final SpMV results.

\begin{figure}[tb]
\centering
\includegraphics[width=0.8\columnwidth]{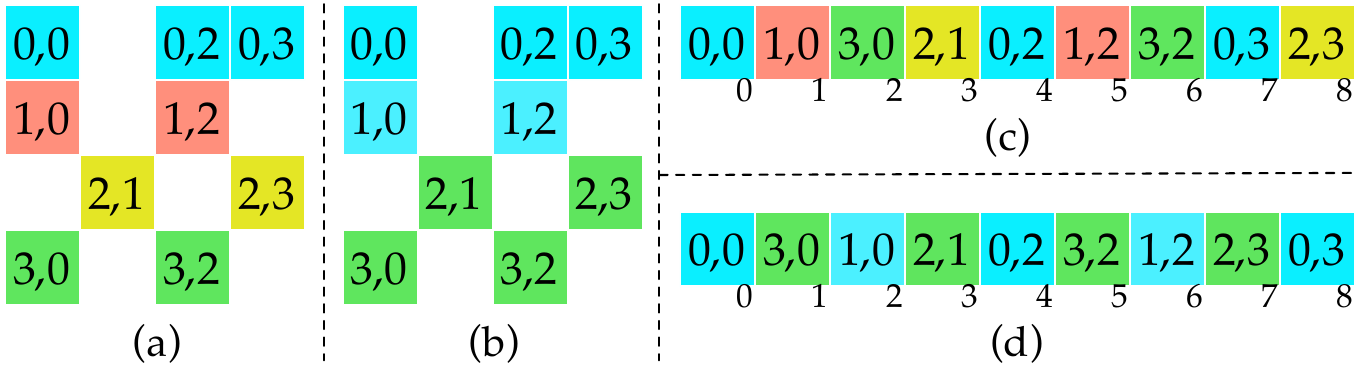}
\vspace{-12pt}
\caption{An example of "coloring" and reordering in Sextans~\cite{song2021sextans}: (a), (c), and
\serpens: (b), (d). We assume the DSP latency is 2.
}
\label{figure:reorder}
\vspace{-12pt}
\end{figure}

\subsection{SpMV Processing Order}
\label{sec:process_order}

Figure\ref{figure:overall_arch} (b) shows the
overall processing order in \serpens.
We partition the dense vector $\vec{x}$
into segments. The segment length $W=8192$.
In the processing, we stream in one $\vec{x}$ segment and store it in BRAMs.
Then we stream in sparse $\mathbf{A}$ elements associated to the $\vec{x}$ segment.
For the accumulation of $\mathbf{A}\times\vec{x}$, we use URAMs as accumulation buffers.
After we finish the processing on
one $\vec{x}$ segment, we iterate
on next segment. With the partition
and SpMV processing order, the benefits include:
(i) \serpens does not issue any random access to off-chip memory. All the off-chip memory accessing is sequential. Thus, \serpens can fully utilize the off-chip memory bandwidth and
amortise the latency of off-chip memory accessing.
(ii) There are two kind of random accesses in SpMV: (1) irregularly reading elements from the dense vector $\vec{x}$ and 
(2) accumulating on vector $\vec{y}$. 
Now, \serpens limits the two random accessing on chip. With the optimization of II=1 pipeline, we can achieve a high processing throughput.
(iii) \serpens will
read/write any of the vectors and the sparse matrix only once without any duplicate reading/writing. Thus, \serpens
minimizes the off-chip communication.

\subsection{Memory-Centric Processing Engines}
HBM provides massive memory channels to users.
However, the switching/crossing between
HBM channels requires special handling to achieve
high memory throughput~\cite{choi2021hbm}.
\serpens avoid one module to access cross
multiple HBM channels by
the memory-centric processing engines
as shown in Figure\ref{figure:overall_arch} (c).
\serpens distributes
the sparse elements from one channel to 8 PEs.
We implement the broadcasting of the
dense vector segment from one channel to all 
PEs using a chain topology~\cite{cong2018latte}
to achieve a higher frequency~\cite{guo2020analysis}.
There are four copies of the the dense vector
$\vec{x}$ segment stored in BRAMs. Since each BRAM has two ports, we share one BRAM with two PEs
to save half of the BRAM usage. 
There
is no bank conflict when the 8 PEs fetch
dense elements from different BRAM addresses.
We make URAM address for each PE disjoint 
to avoid the URAM bank conflict caused by the accessing
from multiple PEs.

The memory-centric PEs also enable rapid
scalability. Users can easily customize \serpens
according to their need on various memory
channels and bandwidths. We show the performance
of \serpens when we scale up channel allocation
for the sparse matrix
from 16 to 24 in
Sec.~\ref{sec:scalable}.

\subsection{Index Coalescing \& Non-Zero Reordering}

\begin{table}[tb]
\caption{The design parameters of \serpens accelerator.}
  \label{table:parameters}
  \vspace{-9pt}
  \centering
  \footnotesize
  \begin{tabular}{cccc}
    \hline
    \multicolumn{4}{c}{\textbf{Architecture}}\\
    \hline
    HBM Channels & PEs/Channel &
    BRAM18Ks/PE
    & URAMs/PE \\
    $H_A=16/24$ & 8 & 128 & $U=3$ 
     \\
    \hline
  \end{tabular}
  \\
  \vspace{3pt}
  \begin{tabular}{cccc}
    \hline
    \multicolumn{4}{c}{\textbf{Bit-Width}}\\
    \hline
    Memory Bus & Data  & Index 
    & Instruction \\
    512 & 32(float) & 32(row+col.) & 32 
     \\
    \hline
  \end{tabular}
  \vspace{-18pt}
  
\end{table}

Index coalescing is a micro-architecture level 
optimization we apply to improve URAM utilization.
The minimum bit width of a URAM configuration is 72.
It is a waste to store a 32-bit float (FP32) value to one URAM
address (entry). Thus, we coalesce two values whose
destination row indices are consecutive into
one common URAM address. In the reordering of
non-zeros (sparse elements), besides read-after-write
(RAW) conflict~\cite{song2021sextans}, we also need to handle
the URAM access conflicts caused by coalesced indices.

Figure\ref{figure:reorder} compares
the non-zero reordering 
in Sextans~\cite{song2021sextans} and in \serpens.
We assume the latency $T$ of DSP accumulation on a 
float is 2 cycles. We view the recording in
Sextans~\cite{song2021sextans} as a process of two 
steps -- coloring and reordering. The coloring step
colors elements in the same row with the same color
shown in Figure~\ref{figure:reorder} (a).
In the recording step, we ensure 
none of the same color elements are in any of 
the $T$-cycle windows (Figure~\ref{figure:reorder} (c)). 
After we applied
the index coalescing, we store two elements of 
consecutive row indices to the same URAM address.
To avoid URAM address conflicts, we only need to 
color elements of two consecutive rows with
the same color shown in Figure~\ref{figure:reorder} (b) 
and then apply the same reordering rule for \serpens. 
Figure~\ref{figure:reorder} (d) illustrates the reordered 
non-zeros taking both RAW and index coalescing 
conflicts into consideration.
We summarize the design parameters of \serpens accelerator
in Table~\ref{table:parameters}.

\subsection{Resource \& Performance Analysis}

We estimate \serpens BRAM/URAM
consumption and cycle count when
achieving an II=1 pipeline.

\subsubsection{BRAM Consumption}
When streaming in the dense vector and storing the
vector to BRAMs, one 512-bit block contains
16 FP32 values. The bitwidth of a BRAM18K is 18, 
so we need 2 BRAM18Ks to store one FP32. For the whole
512 bits, we require 32 BRAM18Ks. Since one BRAM18K has
two ports, we reduce the BRAM18K number to 16.
When streaming in sparse elements,
for each memory channel, we 
dispatch 8 sparse elements to 8 PEs per cycle.
Thus we need $16\times8=128$ BRAM18Ks. 
Because we share one BRAM with two PEs in \serpens,
the actual number of BRAM18Ks per channel is 64.
We assume there are $H_A$ HBM 
memory channels allocated
to the sparse matrix $\mathbf{A}$. In total, we require
$64\cdot H_A$ BRAM18Ks. One BRAM contains
two BRAM18Ks on Xilinx FPGAs. Thus, the number of BRAMs is:
\vspace{-3pt}
\begin{equation}
\text{\#BRAMs} = 32\cdot H_A.
\vspace{-3pt}
\end{equation}

\subsubsection{URAM Consumption}
Assuming that we assign $U$ URAMs to each PE. There are
$8\cdot H_A$ PEs in total, because 
URAMs are disjoint for different PEs, the total number of
URAMs is:
\vspace{-3pt}
\begin{equation}
\text{\#URAMs} = 8\cdot H_A\cdot U.
\vspace{-3pt}
\end{equation}
Assuming the depth of a URAM configured by a wdith of 72 bits is $D$, with the index coalescing in \serpens,
the on-chip accumulation row depth is:
\vspace{-3pt}
\begin{equation}
\text{\#Row Depth} = 16\cdot H_A\cdot U\cdot D.
\vspace{-3pt}
\end{equation}

\subsubsection{Cycle Count}

We assume the row number, column number, and number
of non-zeros (sparse elements) of the sparse matrix is
$M$, $K$, and NNZ, respectively.

Because we allocate one memory channel to
the dense vector $\vec{x}$,
the cycle count of streaming in
vector $\vec{x}$ is $K/16$.
\serpens performs the
streaming in the dense input vector $\vec{y}$ and
the streaming the out dense output vector $\vec{y}$
in parallel, thus the cycle count on the two 
dense $\vec{y}$ vectors
is $M/16$.

In the computation, at each cycle,
one PE processes 8 sparse elements.
Because there are $8\cdot H_A$ PEs in
total, the cycle count for processing 
the sparse elements is $\text{NNZ} / (8\cdot H_A)$. 

We add up the streaming and computation cycle counts to 
obtain the overall cycle count:
\vspace{-3pt}
\begin{equation}
\text{\#Cycle} = (M + K) / 16 + \text{NNZ} / (8\cdot H_A).
\vspace{-3pt}
\end{equation}

%% file: texfiles/sec_evaluation.tex
\subsection{Evaluation Setup}

\subsubsection{The Evaluated Accelerators}

We evaluate the SpMV routine on 
\serpens and two FPGA-related
accelerators -- Sextans~\cite{song2021sextans}, 
GraphLily~\cite{hu2021graphlily}, 
and an
Nvidia Tesla K80 GPU. Table~\ref{table:platforms}
lists the frequency, memory bandwidth and power of the four evaluated accelerators.

We describe \serpens accelerator in Xilinx high level synthesis (HLS)
C++ and prototype with
Vitis 2020.2. 
For Sextans, we utilize the open-sourced code and prototype
with Vitis 2020.2. 
For GraphLily, we obtain the 
open-sourced bitstream (.xclbin).
We run the three FPGA accelerators 
on a Xilinx Alveo U280 FPGA board
then measure the execution
time by Xilinx Run Time and the power 
consumption by {\tt xbutil}. 
To perform SpMV on GPU,
we use CuSPARSE~\cite{naumov2010cusparse} routine {\tt csrmv} with CUDA 10.1. We measure the GPU execution time
by {\tt cudaEventElapsedTime} and power consumption
by {\tt nvidia-smi}.
We amortize 
the execution time by 100 runs.

GraphLily employs 19 HBM channels and 1 DDR4 channel, translating to 285 GB/s memory bandwidth. Sextans uses 29 HBM channels and the bandwidth is 417 GB/s. \serpens-A16 uses 19 HBM channels for a bandwidth of
273 GB/s.

\subsubsection{The Evaluated Matrices}

\begin{table}[tb]
\caption{The specification of the evaluated accelerators.}
  \label{table:platforms}
  \vspace{-9pt}
  \centering
  \footnotesize
  \begin{tabular}{c|cccc}
    \hline
    & \textbf{Sextans}~\cite{song2021sextans} & \textbf{GraphLily}~\cite{hu2021graphlily} & \textbf{\serpens} & \textbf{Tesla K80} \\
    \hline
    Frequency & 197 MHz & 166 MHz & 223 MHz & 562 MHz \\
    Bandwidth & $^\&$417 GB/s & $^\&$285 GB/s & $^\&$273 GB/s & $^\#$480 GB/s \\
    Power & 52 W & 43 W & 48 W & 130 W\\
    \hline
  \end{tabular}\\
  $^\&$ Utilized bandwidth, $^\#$ maximum bandwidth.
  \vspace{-9pt}
  
\end{table}

\begin{table}[tb]
\caption{The specification of evaluated matrices.}
\vspace{-9pt}
  \label{table:matrices}
  \centering
  \footnotesize
  \begin{tabular}{cccc}
    \hline
    \multicolumn{4}{c}{Twelve Large Matrices/Graphs}\\
    \hline
    \textbf{ID} & \textbf{Matrix} & \textbf{\#Vertices} & \textbf{\#Edges} \\
    \hline
    G1 & {\tt googleplus}~\cite{snapnets} & 108 K & 13.7 M \\
    G2 & {\tt crankseg\_2}~\cite{davis2011university} & 63.8 K & 14.1 M  \\
    G3 & {\tt Si41Ge41H72}~\cite{davis2011university} & 186 K & 15.0 M \\
    G4 & {\tt  TSOPF\_RS\_b2383}~\cite{davis2011university} & 38.1 K & 16.2 M \\
    G5 & {\tt ML\_Laplace}~\cite{davis2011university} & 377 K & 27.6 M \\
    G6 & {\tt mouse\_gene}~\cite{davis2011university} & 45.1 K & 29.0 M \\
    G7 & {\tt soc\_pokec}~\cite{snapnets} & 1.63 M & 30.6 M  \\
    G8 & {\tt coPapersCiteseer}~\cite{davis2011university} & 434 K & 21.1 M \\
    G9 & {\tt PFlow\_742}~\cite{davis2011university} & 743 K & 37.1 M \\
    G10 & {\tt ogbl\_ppa}~\cite{hu2020open} & 576 K & 42.5 M \\
    G11 & {\tt hollywood}~\cite{snapnets} & 1.07 M & 113 M \\
    G12 & {\tt ogbn\_products}~\cite{hu2020open} & 2.45 M & 124 M \\
    \hline
  \end{tabular}\\ \vspace{3pt}
  \footnotesize
  \begin{tabular}{cc|cc}
    \hline
    \multicolumn{4}{c}{SuiteSparse~\cite{davis2011university} Matrices}\\
    \hline
    \textbf{Number of Matrices}  & 2,519 
    & \textbf{NNZ}  & 1,000 -- 89,306,020 \\
    \textbf{Row/column}  & 24 -- 2,999,349
    & \textbf{Density}  & 8.75E-7 -- 1 \\
    \hline
  \end{tabular}
  \vspace{-9pt}
\end{table}

\begin{table*}[tb]
  \caption{Performance of Sextans~\cite{song2021sextans}, GraphLily~\cite{hu2021graphlily}, and \serpens on twelve large matrices/graphs. The improvement is the ratio of a performance metric of \serpens compared to that of GraphLily.}
  \label{table:perf}
  \vspace{-9pt}
  \footnotesize
  \centering
  \begin{tabular}{lrccccccccccccc}
    \hline
    & & G1 & G2 & G3 & G4 & G5 & G6 & G7 & G8 & G9 & G10 & G11 & G12 & \textbf{GMN} \\
    \hline
    \multirow{3}{6mm}{Execution Time: ms} & \textbf{Sextans} &
    3.06 & 1.38 & 1.64 & 1.36 & 2.73 & 2.72 & -- & 3.58& -- & -- & -- & -- & 2.20
    \\
    & \textbf{GraphLily} &
    1.73 & 1.47 & 1.85 & 1.57 & 2.96 & 2.80 & 7.04 & 3.63 & 4.52 & 4.59 & 12.4 & 18.6 & 3.74
    \\
    & \textbf{\serpens-A16} &
    1.87 & 0.930 & 0.853 & 0.730 & 1.37 & 1.37 & 4.52 & 2.09 & 2.05 & 2.04 & 6.20 & 6.32 & 1.96
    \\

    \hline
    \multirow{3}{12mm}{Throughput: GFLOP/s} & \textbf{Sextans} &
    9.01 & 20.60 & 18.55 & 23.81 & 20.47 & 21.33 & - & 18.14 & - & - & - & - & 18.15
    \\
    & \textbf{GraphLily} &
    15.96 & 19.36 & 16.44 & 20.64 & 18.87 & 20.69 & 9.17 & 17.90 & 16.75 & 18.74 & 18.36 & 13.60 & 16.86
    \\
    & \textbf{\serpens-A16} &
    14.71 & 30.56 & 35.62 & 44.39 & 40.75 & 42.26 & 14.29 & 31.06 & 37.01 & 42.26 & 36.70 & 39.90 & 32.21
    \\
    
    
    \hline
    \multirow{4}{12mm}{Throughput: MTEPS} & \textbf{Sextans} &
    4,470 & 10,255 & 9,162 & 11,878 & 10,099 & 10,651 & -- & 8,951 & -- & -- & -- & -- & 9,005
    \\
    & \textbf{GraphLily} &
    7,920 & 9,639 & 8,117 & 10,296 & 9,305 & 10,331 & 4,352 & 8,828 & 8,212 & 9,243 & 9,094 & 6,668 & 8,310
    \\
    & \textbf{\serpens-A16} &
    7,300 & 15,214 & 17,594  & 22,144 & 20,099 & 21,098 & 6,782 & 15,324 & 18,142 & 20,847 & 18,176 & 19,565 & 15,876
    \\
    & \textbf{Improvement} &
    0.922$\times$ & 1.58$\times$ & 2.17$\times$ & 2.15$\times$ & 2.16$\times$ & 2.04$\times$ & 1.56$\times$ & 1.74$\times$ & 2.21$\times$ & 2.26$\times$ & 2.00$\times$ & 2.93$\times$ & 1.91$\times$
    \\
    \hline

    \multirow{4}{15mm}{Bandwidth Efficiency: MTEPS/(GB/s)} & \textbf{Sextans} &
    10.7 & 24.6 & 22.0 & 28.5 & 24.2 & 25.5 & -- & 21.5 & -- & -- & -- & -- & 21.6
    \\
    & \textbf{GraphLily} &
    27.8 & 33.8 & 28.5 & 36.1 & 32.7 & 36.2 & 15.3 & 31.0 & 28.8 & 32.4 & 31.9 & 23.4 & 29.2
    \\
    & \textbf{\serpens-A16} &
    26.7 & 
55.7 & 
64.4 & 
81.1 & 
73.6 & 
77.3 & 
24.8 & 
56.1 & 
66.5 & 
76.4 & 
66.6 & 
71.7 & 
58.2
    \\
    & \textbf{Improvement} &
0.962$\times$ &
1.65$\times$ &
2.26$\times$ &
2.25$\times$ &
2.25$\times$ &
2.13$\times$ &
1.63$\times$ &
1.81$\times$ &
2.31$\times$ &
2.35$\times$ &
2.09$\times$ &
3.06$\times$ &
1.99$\times$ 
    \\
    \hline

    \multirow{4}{15mm}{Energy Efficiency: MTEPS/W} & \textbf{Sextans} &
86.0 &
197 &
176 &
228 &
194 &
205 &
--&
172 &
--&
--&
--&
--&
173
    \\
    & \textbf{GraphLily} &
184 &
224 &
189 &
239 &
216 &
240 &
101 &
205 &
191 &
215 &
211 &
155 &
193
    \\
    & \textbf{\serpens-A16} &
152 &
317 &
367 &
461 &
419 &
440 &
141 &
319 &
378 &
434 &
379 &
408 &
331
    \\
    & \textbf{Improvement} &
0.826$\times$ &
1.41$\times$ &
1.94$\times$ &
1.93$\times$ &
1.94$\times$ &
1.83$\times$ &
1.40$\times$ &
1.56$\times$ &
1.98$\times$ &
2.02$\times$ &
1.79$\times$ &
2.63$\times$ &
1.71$\times$ 

    \\
    \hline
    
  \end{tabular}
  \vspace{-9pt}
\end{table*}

For the comparison of \serpens
with Sextans and GraphLily,
we evaluate them on 12 large matrices/graphs
which are selected from SNAP~\cite{snapnets}, OGB~\cite{hu2020open}, and SuiteSparse~\cite{davis2011university}.
The number of vertices (rows) ranges from 
45K to 2.45M and the number of edges (non-zeros) 
can be as high as 124M. For the 
comparison of \serpens with K80, we evaluate on 2,519 sparse matrices whose number of non-zeros (NNZ) is greater than 1,000 and less than 100M from
SuiteSparse. The geomean density of the
evaluated SuiteSparse matrices is 1.4E-3.
Table~\ref{table:matrices}
shows the specifications of the evaluated
matrices. 
We evaluate single floating-point 
SpMV.
We compare the execution time (ms), throughput in million 
traversed edges per second (MTEPS),
bandwidth efficiency defined as
(throughput)/(memory bandwidth), 
and energy efficiency
defined as
(throughput)/(power consumption)
of the three FPGA accelerators.
We set N=8 (the minimal supported N) for Sextans~\cite{song2021sextans}
to obtain SpMV results.
We run
GraphLily~\cite{hu2021graphlily}
on SpMV mode.

\subsection{Comparison with Related Accelerators}

Table~\ref{table:perf} shows the execution time, throughput, 
bandwidth and energy efficiency
of the three FPGA accelerators.
Sextans~\cite{song2021sextans} is an HBM-FPGA SpMM accelerator. An SpMV accelerator is able to
switch to process SpMM and vice versa in functionality.
However, 
their designs are different and customized for
the performance of tow different kernels -- SpMV/SpMM.
We use Table~\ref{table:comparetosextans} to compare \serpens and Sextans~\cite{song2021sextans}.
We use the matrix
{\tt TSOPF\_RS\_b2383\_c1} from SuiteSparse~\cite{davis2011university} to illustrate
the difference. The SpMM(N=16) latencies
of \serpens (running 16 SpMVs) and Sextans are 8.56 ms and 2.87 ms, respectively, and the SpMV
latencies
of \serpens and Sextans are 0.535 ms and 1.44 ms, respectively. We got lower performance if we
use an SpMM accelerator for perform SpMV and 
vice versa. The different customization in the
accelerators lead to their performance expertise.
For memory channel allocation for the matrices,
since the vector size is significantly smaller than
the sparse matrix size, thus \serpens needs to allocate
one channel for a vector and allocate
memory channels for the sparse matrix. So \serpens performs better than Sextans for SpMV.
However, the dense element sharing helps Sextans
perform better than \serpens for SpMM.
GraphLily is an FPGA overlay which is able to support
a few graph kernels that can be executed in
a BLAS processing model. 
GraphLily supports generalized multiplication and 
generalized reduction. For example, GraphLily
can configure a generalized multiplication as
one of (1) algebraic multiplication, (2)
algebraic addition, (3) logic AND, and (4) zero output.
In the processing of SpMV where GraphLily
configures a generalized
multiplication as an algebraic multiplication, 
the hardware resource for the other two
operations is idle. Thus, GraphLily lacks deeper 
specialization for SpMV and the SpMV
execution time of GraphLily is larger than the
execution time of \serpens.

\begin{table}[tb]
\caption{Comparisons of \serpens, Sextans~\cite{song2021sextans}, and GraphLily~\cite{hu2021graphlily}.}
  \label{table:comparetosextans}
  \vspace{-9pt}
  \centering
  \scriptsize
  \begin{tabular}{rcccc}
    \hline
    & Kernel & \#Ch. - Sparse A & \#Ch. - Dense B/C(X/Y) & \#Ch. - Instr.\\
    \hline
    \textbf{\serpens} & SpMV & 16/24 & 1/1 & 1 \\
    \textbf{Sextans} & SpMM & 8 & 4/8 & 1\\
    \textbf{GraphLily} & Graph & 16 & 1/1 & -\\
    \hline
    \hline
    & OoO NZ & Sharing Sparse A & Index Coalescing & Perf - SpMV/SpMM\\
    \hline
    \textbf{\serpens} & Yes & No & Yes & High/Low \\
    \textbf{Sextans} & Yes & Yes & No & Low/High\\
    \textbf{GraphLily} & No & No & No & -/-\\
    \hline
  \end{tabular}
  \vspace{-6pt}
  
\end{table}

\begin{table}[tb]
  \caption{Resource utilization of Sextans, GraphLily, and \serpens-A16 on a Xilinx U280 FPGA board.}
  \label{table:ultilization}
  \vspace{-9pt}
  \footnotesize
  \centering
  \begin{tabular}{p{10mm}ccccc}
    \hline
    & LUT & FF & DSP & BRAM & URAM \\
    \hline
    \textbf{Sextans}
    & 331K(29\%) & 594K(25\%) & 3233(36\%) & 1238(68\%) & 768(80\%) \\
    \textbf{GraphLily}
    & 390K(35\%) & 493K(21\%) & 723(8\%) & 417(24\%) & 512(53\%) \\
    \textbf{\serpens}
    & 173K(15\%) & 327K(14\%) & 720(8\%) & 655(36\%) & 384(40\%) \\
    \hline
  \end{tabular}
  \vspace{-12pt}
\end{table}

\subsubsection{Execution Time}
Sextans is not able to support Matrix
G7 and G9 -- G12 directly on the hardware.
For the other matrices, 
the execution time of \serpens
is less than the execution time 
of Sextans.
For the comparison with GraphLily, \serpens is slightly slower 
(1.87 ms v.s. 1.73 ms) on
G1 but faster than
GraphLily on the other 11 matrices.

\subsubsection{Throughput} We use (NNZ)/(execution time) to calculate the throughput
(MTEPS). 
The throughput directly corresponds to the execution time. A shorter execution time
leads to a higher throughput.
The maximum throughput achieved by GraphLily is 10,331 MTEPS
while the 
maximum throughput achieved by \serpens is 22,144 MTEPS.
For the geometric throughput,
GraphLily and \serpens achieve
a geomean throughput of 8,310 
MTEPS and 15,876 MTEPS respectively, leading to a
1.91$\times$ throughput improvement of \serpens over GraphLily.

\subsubsection{Bandwidth Efficiency}
On the same graph/matrix, the 
bandwidth efficiency is determined 
by the execution time and the accelerator's 
memory bandwidth. GraphLily's 
bandwidth is higher than \serpens' bandwidth 
(285 GB/s v.s. 273 GB/s). With a faster execution time,
\serpens achieves a geomean bandwidth efficiency of 
58.2 MTEPS / (GB/s), 1.99$\times$ compared with 
 GraphLily's bandwidth efficiency. The highest
 bandwidth efficiency achieved by \serpens is 
81.1 MTEPS / (GB/s) on G4.

\subsubsection{Energy Efficiency.}
Some hardware resource of GraphLiLy overlay
may be idle when performing one specific 
graph kernel, so the power consumption
of GraphLiLy is lower than that of \serpens
(43 W v.s. 48 W). However, \serpens is
$1.91\times$ faster than GraphLiLy, leading
to a $1.71\times$ energy efficiency improvement.

\subsubsection{Resource Utilization}

Table~\ref{table:ultilization} lists the
FPGA resource utilization of the three 
accelerators on the same U280 board.
Sextans requires the highest resource utilization
because Sextans needs to compute on the dense
$\mathbf{B}$ matrix which is larger than 
the $\vec{x}$ vector 
in SpMV. 
In contrast to GraphLily,
\serpens consumes less LUT, FF, DSP, and URAM.
Because \serpens is customized specifically for SpMV,
it does not need the extra FPGA resource
that GraphLily requires for
its generalized operations.
However, \serpens consumes more BRAMs than 
GraphLily, because \serpens explicitly 
deploys more BRAMs to access
on-chip memory in parallel.

\subsubsection{Other SpMV Accelerators.} We compare \serpens with two other real-execution SpMV accelerators \cite{sadi2019efficient} and SparseP~\cite{giannoula2022sparsep} in Table~\ref{table:otherspmv}. \cite{sadi2019efficient} is based
on an FPGA and SparseP~\cite{giannoula2022sparsep} is based
on a real PIM system. \serpens-A24 has the highest peak performance and \serpens-A16 performs better and has lower
memory bandwidth than both
\cite{sadi2019efficient} and \cite{giannoula2022sparsep}.

\begin{table}[tb]
  \caption{Comparison with other SpMV accelerators.}
  \label{table:otherspmv}
  \vspace{-9pt}
  \footnotesize
  \centering
  \begin{tabular}{rcc}
    \hline
    & Bandwidth & Peak Performance \\
    \hline
    \textbf{\serpens-A16} & 273 GB/s & 44.2 GFLOP/s \\
    \textbf{\serpens-A24} & 388 GB/s & 60.4 GFLOP/s \\
    \textbf{\cite{du2022high}} & 258 GB/s & 25.0 GFLOP/s \\
    \textbf{\cite{sadi2019efficient}} & 357 GB/s & 34.0 GFLOP/s \\
    \textbf{SparseP~\cite{giannoula2022sparsep}} & 1770 GB/s & 4.66 GFLOP/s \\
    \hline
  \end{tabular}
  \vspace{-6pt}
\end{table}

\subsection{Comparison with K80 GPU}
\begin{figure}[tb]
\centering
\includegraphics[width=0.9\columnwidth]{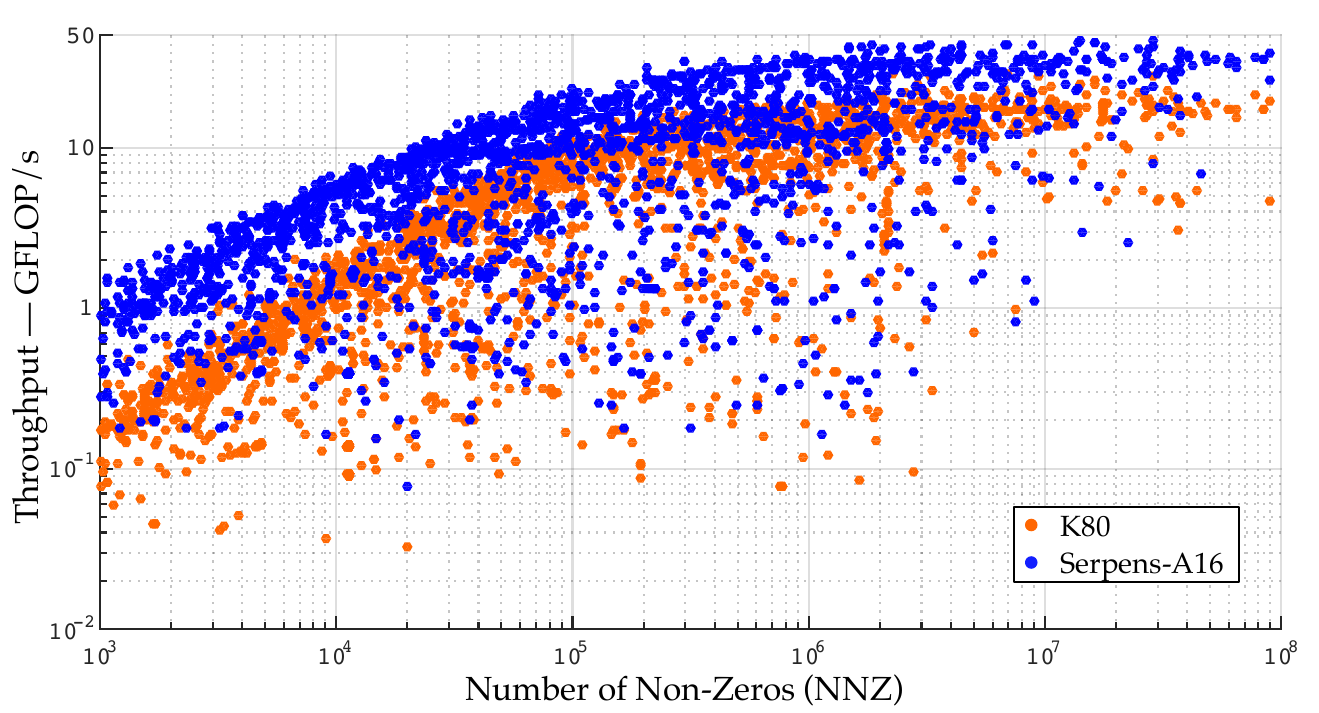}
\vspace{-9pt}
\caption{SpMV throughput (in GFLOP/s) of K80 and \serpens plotted with increasing NNZ. 
}
\label{figure:mteps}
\vspace{-6pt}
\end{figure}

We compare \serpens-A16 with K80 GPU on a wide
range of 2,519 sparse matrices from
SuiteSparse~\cite{davis2011university} to 
demonstrate the performance of \serpens
as a general-purpose accelerator 
in a data center.
K80 is a more powerful accelerator
than \serpens in terms of frequency
and bandwidth as shown in Table~\ref{table:platforms}.

We plot the SpMV throughputs of K80
and \serpens in Figure~\ref{figure:mteps}.
\serpens achieves higher throughput on almost
all matrices than K80. 
The maximum throughputs of K80 and \serpens are 
46.43~GFLOP/s (14,521~MTEPS) and
29.12~GFLOP/s (23,158~MTEPS)
respectively.
The geomean throughput of \serpens
compared to
K80 is $2.31\times$. 
For the geomean bandwidth efficiency,
K80 achieves 2.10 MTEPS /(GB/s). 
With a geomean bandwidth efficiency of
8.52 MTEPS/(GB/s),
\serpens outperforms K80 by $4.06\times$.
For the geomean energy efficiency,
K80 achieves 7.75 MTEPS/W. 
\serpens has a 
geomean energy efficiency of
48.4 MTEPS/W ($6.25\times$ better).

\subsection{Scalability}
\label{sec:scalable}
We scale up HBM channel allocation
from 16 to 24
to further boost 
performance. Vanilla Vitis failed
place and route because of the congestion
caused by heavy HBM channel usage.
With the aid of TAPA~\cite{chi2021extending} and
Autobridge~\cite{guo2021autobridge},
we successfully place and route the 24 HBM channel
version, resulting in
270 MHz frequency.
We compare \serpens-A24 with GraphLily~\cite{hu2021graphlily} in Table~\ref{table:A24}.
\serpens-A24 achieves up to
60.55~GFLOP/s (30,204~MTEPS)
and a throughput of up to 
3.79$\times$ improvement 
over GraphLily.

\begin{table}[tb]
  \caption{The SpMV throughput (GFLOP/s) of the 24 HBM channel version \serpens and improvement over  GraphLily.}
  \label{table:A24}
  \vspace{-9pt}
  \footnotesize
  \centering
  \begin{tabular}{rcccccc}
    \hline
    & G1 & G2 & G3 & G4 & G5 & G6 \\
    \hline
    \textbf{\serpens-A24} &  
15.33 &
36.05 &
45.07 &
60.55 &
52.30 &
57.96
\\
    \textbf{Improvement} &
0.960$\times$ &
1.86$\times$ &
2.74$\times$ &
2.93$\times$ &
2.77$\times$ &
2.80$\times$ 
\\
    \hline
    & G7 & G8 & G9 & G10 & G11 & G12 \\
    \hline
    \textbf{\serpens-A24} & 
18.34 &
36.47 &
46.86 &
56.11 &
45.08 &
51.56
\\
    \textbf{Improvement} &
2.00$\times$ &
2.04$\times$ &
2.80$\times$ &
3.00$\times$ &
2.46$\times$ &
3.79$\times$
\\
    \hline
  \end{tabular}
  \vspace{-9pt}
\end{table}

%% file: texfiles/sec_conclusion.tex
We present \serpens, an HBM based accelerator for SpMV acceleration.
\serpens is a general-purpose design
which supports an arbitrary SpMV.
We design memory-centric processing engines in \serpens for full utilization
of memory bandwidth and the scalability 
with memory channels.
We improve URAM utilization
for vector storage by index coalescing, and the index coalescing is integrated
with non-zero recording. 
In the evaluation, we compare
\serpens with two related FPGA accelerators Sextans~\cite{song2021sextans} and
GraphLily~\cite{hu2021graphlily}.
\serpens outperforms the latest accelerators GraphLiLy and Sextans by
$1.91\times$ and $1.76\times$, respectively, 
in terms of geomean throughput.
For the comparison of \serpens with 
K80 GPU on SuiteSparse~\cite{davis2011university}, \serpens achieves $2.10\times$ higher throughput.
We scale up \serpens to support 24 HBM 
channels for the spares matrix. 
After scaling up to 24 HBM channels,
\serpens achieves a throughput of up to
60.55~GFLOP/s (30,204~MTEPS) and up to 
3.79$\times$ over GraphLily.